\documentclass{jfm}
\usepackage{natbib}
\usepackage[dvips]{graphicx}
\usepackage{amsmath}
\usepackage{amssymb}
\usepackage{subfigure}
\usepackage{color}

\newcommand{\EMPTY}[1]{{}}

\def\maxsize{.49\textwidth}
\def\vissize{.46\textwidth}

\newcommand{\const}{{\rm const}}

\newcommand{\beq}{\begin{equation}}
\newcommand{\eeq}{\end{equation}}
\newcommand{\bea}{\begin{eqnarray}}
\newcommand{\eea}{\end{eqnarray}}

\newcommand{\dd}{\partial} 
\newcommand{\diffr}[1]{\frac{\dd #1}{\dd r}}

\newcommand{\difft}[1]{\frac{\dd #1}{\dd t}}

\newcommand{\lt}{\left}
\newcommand{\rt}{\right}

\newcommand{\secref}[1]{\S\ref{#1}}

\newcommand{\Eqref}[1]{Equation~(\ref{#1})}
\renewcommand{\eqref}[1]{equation~(\ref{#1})}

\newcommand{\eqsand}[2]{equations~(\ref{#1}) and~(\ref{#2})}

\newcommand{\eqsdash}[2]{equations~(\ref{#1})--(\ref{#2})}

\newcommand{\exref}[1]{(\ref{#1})}

\newcommand{\Figref}[2]{Figure~\ref{#1}({\em #2})}
\newcommand{\figref}[2]{figure~\ref{#1}({\em #2})}

\newcommand{\pluses}{+++}
\newcommand{\solid}{{\color{magenta}---}}
\newlength{\oldfboxrule}
\newlength{\oldfboxsep}
\oldfboxrule=\fboxrule
\oldfboxsep=\fboxsep
\newcommand{\boldsolid}{{\fboxrule=2pt\fboxsep=0pt\tiny\raise1pt\hbox{\colorbox{black}{------}}\fboxrule=\oldfboxrule\fboxsep=\oldfboxsep}}
\newcommand{\dotted}{$\cdot\cdot\cdot$}
\newcommand{\dashed}{{\color{red}$---$}}
\newcommand{\dashdotted}{{\color{blue}$-\cdot-$}}
\newcommand{\dashtridot}{$-\cdot\cdot\cdot-$}

\newcommand{\smalll}{L}

\newcommand{\avg}[1]{\left<#1\right>}

\renewcommand{\vec}[1]{\boldsymbol{#1}}
\newcommand{\vdel}{\vec{\nabla}}
\newcommand{\vx}{\vec{x}}
\newcommand{\vr}{\vec{r}}
\newcommand{\rr}{\hat r} 
\newcommand{\vrr}{\vec{\rr}}
\newcommand{\vu}{\vec{u}}
\newcommand{\dvu}{\delta\vu}
\newcommand{\du}{\delta u} 
\newcommand{\ul}{u_\smalll} 
\newcommand{\dul}{\delta\ul}

\newcommand{\vB}{\vec{B}}
\newcommand{\dvB}{\delta\vB}
\newcommand{\dB}{\delta B} 
\newcommand{\Bl}{B_\smalll} 
\newcommand{\dBl}{\delta\Bl}

\newcommand{\vz}{\vec{z}}
\newcommand{\dvz}{\delta\vz}
\newcommand{\dz}{\delta z} 
\newcommand{\zl}{z_\smalll} 
\newcommand{\dzl}{\delta\zl}
\newcommand{\vf}{\vec{f}}
\newcommand{\fl}{f_\smalll} 

\newcommand{\eps}{\epsilon}

\newcommand{\urms}{u_{\rm rms}}
\newcommand{\Brms}{B_{\rm rms}}
\newcommand{\lf}{\ell_0}
\newcommand{\lstar}{\ell_*}
\newcommand{\lres}{\ell_\eta}

\newcommand{\lvisc}{\ell_\nu}
\newcommand{\lpar}{\ell_\parallel}

\newcommand{\Pm}{\mbox{Pm}}
\renewcommand{\Re}{\mbox{Re}}
\newcommand{\Rm}{\mbox{Rm}}

\nocite{*}

\begin{document}

\title[Exact scaling laws and the local structure of isotropic MHD
turbulence]{Exact scaling laws and the local structure of isotropic 
magnetohydrodynamic turbulence}

\author[T.\ A.\ Yousef, F.\ Rincon and A.\ A.\ Schekochihin]{T.\ls A.\ls Y\ls O\ls U\ls S\ls E\ls F,\ns 
F.\ls R\ls I\ls N\ls C\ls O\ls N \and\!\! 
A.\ls A.\ls S\ls C\ls H\ls E\ls K\ls O\ls C\ls H\ls I\ls H\ls I\ls N}
\affiliation{Department of Applied Mathematics and Theoretical Physics, 
University of Cambridge, Wilberforce Road, Cambridge CB3 0WA, United Kingdom}
%\label{firstpage}

\maketitle

\begin{abstract}
This paper examines the consistency of the exact scaling laws for 
isotropic MHD turbulence 
in numerical simulations with large magnetic Prandtl numbers $\Pm$ 
and with $\Pm=1$. The exact laws are used to elucidate the structure 
of the magnetic and velocity fields. Despite the linear scaling of 
certain third-order correlation functions, 
the situation is not analogous to the case of Kolmogorov turbulence. 
The magnetic field is adequately described by a model of stripy (folded) field with 
direction reversals at the resistive scale. At currently available resolutions, 
the cascade of kinetic energy is short-circuited by the direct 
exchange of energy between the forcing-scale motions and the stripy 
magnetic fields. This nonlocal interaction is the defining feature of 
isotropic MHD turbulence. 

\end{abstract}

\section{Introduction \label{sec:intro}}

What is the structure of the saturated state of isotropic MHD turbulence? 
This is possibly the oldest question in the theory of MHD turbulence 
\citep{batch50} --- the answer to which is fundamentally important 
to our understanding of cosmic magnetism 
\citep[see, e.g, review by][and references therein]{schek06}. 
The problem can be posed in the following way. 
Consider the equations of incompressible MHD: 
\bea
\label{eq:mom}
\difft{\vu} + \vu\cdot\vdel\vu &=& -\vdel p + \vB\cdot\vdel\vB + \nu\Delta\vu + \vf,\\ 
\label{eq:ind}
\difft{\vB} + \vu\cdot\vdel\vB &=& \vB\cdot\vdel\vu + \eta\Delta\vB, 
\eea
where $\vu$ is the velocity, $\vB$ 
the magnetic field, $p$ the total pressure determined by the 
incompressibility condition $\vdel\cdot\vu=0$, 
$\vf$ a body force, $\nu$ viscosity and $\eta$ magnetic 
diffusivity (we use units in which 
$p$ is scaled by $\rho$ and $\vB$ by $\sqrt{4\pi\rho}$, 
where $\rho=\const$ is the density of the medium). 
The evolution equations for the kinetic and magnetic energies~are 
\bea
\label{eq:EK}
{d\over d t}{\avg{u^2}\over2} &=& -\avg{\vB\vB:\vdel\vu} - \nu\avg{|\vdel\vu|^2} + \eps,\\
\label{eq:EM}
{d\over d t}{\avg{B^2}\over2} &=& \avg{\vB\vB:\vdel\vu} - \eta\avg{|\vdel\vB|^2},
\eea
where $\avg{\dots}$ means volume averaging and $\eps=\avg{\vu\cdot\vf}$ is 
the average injected power per unit volume (formally 
speaking, $\eps$ depends on $\vu$ and cannot 
be predicted in advance unless $\vf$ is a white noise in time  
--- in the latter case, $\eps$ is fixed, which makes white-noise forcing 
an attractive modelling choice). 
%\footnote{Formally 
%speaking, $\eps$ depends on $\vu$ and cannot 
%be predicted in advance unless $\vf$ is a white noise. 
%In the latter case, $\eps$ is fixed. This makes white-noise forcing 
%an attractive modeling choice.} 
The forcing acts 
at the system scale $\lf$ and we shall assume that the particular choice 
of $\vf$ does not change the properties of turbulence at 
scales much smaller than $\lf$. In hydrodynamics, 
one often considers decaying, rather than forced, turbulence. 
The structure of the turbulence at small scales is expected 
to be the same for the decaying and forced cases. This appears 
not to be true for MHD, probably because of a tendency 
of the magnetic field to decay towards large-scale force-free 
states (compare, e.g., the 
decaying simulations of \citealt{bisk00} with the forced 
ones of \citealt{schek04} or \citealt{haugen04}). 
%\footnote{In hydrodynamics, 
%one often considers decaying, rather than forced, turbulence. 
%%and the Kolmogorov energy flux is $\eps=-(1/2)d\avg{u^2}/dt$. 
%The structure of the turbulence at small scales is expected 
%to be the same for the decaying and forced cases. This appears 
%not to be true for MHD, probably because of a tendency 
%of the magnetic field to decay towards large-scale force-free 
%states (compare, e.g., the 
%decaying simulations of \citealt{bisk00} with the forced 
%ones of \citealt{schek04} or \citealt{haugen04}).} 
We consider the case in which the velocity field is forced, while 
the magnetic field is not and can receive energy only via 
interaction with the velocity field. If there is dynamo action, 
an initially weak magnetic 
field is amplified by the turbulence until it becomes dynamically 
significant and saturates. 
Certain choices of the forcing, e.g., helical random forcing \citep{brand01}, 
%or a class of non-random stationary forcing 
%functions that give rise to turbulence with a mean flow 
%$\avg{\vu}\neq0$, 
lead to the generation of magnetic field 
at scales larger than the forcing scale $\lf$. In the presence of  
this {\em mean field}, the turbulence at the small scales is 
(globally) anisotropic --- a distinct case that will not be 
discussed here \citep[see review by][]{schek06}. 
Turbulence produced by a 
spatially homogeneous isotropic nonhelical random forcing 
does not generate a mean field, but, at least when the 
magnetic Prandtl number $\Pm=\nu/\eta\ge1$, does lead to 
amplification of small-scale ($r<\lf$) magnetic energy 
\citep[e.g.,][]{schek04,haugen04}, 
a process known as {\em small-scale dynamo}. 
The saturated state of the small-scale dynamo is the fully developed 
isotropic MHD turbulence that is the subject of this 
paper. 

While some level of physical understanding of the small-scale dynamo 
and its saturation does exist (\secref{sec:ssd}), the detailed structure 
of the saturated state is still unknown. 
The existence of exact scaling laws analogous to Kolmogorov's 
$4/5$ and Yaglom's $4/3$ laws (\eqsand{eq:ff}{eq:ft} below)
has occasionally been interpreted as suggesting that MHD turbulence 
has inertial-range scaling analogous to the hydrodynamic case: 
specifically, that the fundamental physical fields in MHD are 
Elsasser variables $\vz^\pm=\vu\pm\vB$, which 
have Kolmogorov scaling (from \eqref{eq:ft}, the increments 
$\dz^\pm(r)\sim r^{1/3}$ for point separations $r$ 
greater than the viscous and resistive scales), and also that  
the magnetic and kinetic energies are in scale-by-scale equipartition, 
$\du(r)\sim\dB(r)$. Obviously, the exact laws by themselves  
do not rigorously imply any of this. The qualitative analogy with 
Kolmogorov turbulence is based on the assumption that interactions 
occur between comparable scales (locality in scale space) 
and that the exact laws therefore describe a scale-by-scale 
energy cascade. 
However, in isotropic MHD turbulence, the interactions 
are not local. An obvious example 
of a nonlocal process is the small-scale dynamo, whereby 
magnetic fields with direction reversals at the resistive 
scale (the folded fields) are generated by the velocity fluctuations 
that have much larger scales (\secref{sec:ssd}). 
Both numerical evidence and physical reasoning suggest that 
the saturated state of this process 
is controlled by the nonlocal interaction between the 
large-scale (forcing-scale) velocity gradients and the folded (small-scale) 
magnetic structures \citep{schek04,alexakis05sat}. 
It is then an interesting 
question how such a saturated state can be consistent with 
the exact scalings mandated by the exact laws. 
Since these scalings represent the only rigorously established 
constraints on the statistics of the isotropic MHD turbulence, 
understanding how they are satisfied is a way to 
learn something about the structure of the turbulence. 
 
The purpose of this paper is to give 
an interpretation of the exact laws in various scale ranges, 
establish that the laws hold in numerically simulated MHD turbulence, 
and determine the extent to which the 
simulations access the desired asymptotic regimes. 
The plan of further developments is as follows. 
The derivation of the exact laws is reviewed in \secref{sec:exact}. 
A theoretical discussion of the generation of stripy (folded) 
magnetic fields and the structure of the turbulence at subviscous 
scales is given in \secref{sec:theory}. 
Numerical tests are reported in \secref{sec:num}. 
The concluding section, \secref{sec:inertial}, discusses 
isotropic MHD turbulence in the inertial range and 
the unresolved issues. 

\section{Exact scaling laws for isotropic MHD turbulence \label{sec:exact}}

The procedure for obtaining exact scaling laws is standard. 
Consider two points, $\vx$ and $\vx'$. Denote $\vu=\vu(\vx)$, 
$\vu'=\vu(\vx')$, $\dvu=\vu'-\vu$. Use \eqref{eq:mom} 
taken at points $\vx$ and $\vx'$ to derive an evolution equation 
for the correlation tensor $\avg{u_i u_j'}$. 
This equation contains third-order correlation tensors 
$\avg{u_i u_j u_k'}$ and $\avg{B_i B_j u_k'}$. 
Because of the assumed 
spatial homogeneity, all two-point correlation tensors 
depend only on $\vr=\vx'-\vx$. We denote the projections of 
$\vu$ and $\vB$ on $\vr$ by subscript $_\smalll$ 
(for ``longitudinal''), e.g., 
$\dul=\dvu\cdot\vrr$, where $\vrr=\vr/r$. 
Assuming isotropy, all correlation 
tensors can be expressed in terms of a set of scalar 
functions that depend only on $r=|\vr|$. 
%Thus, for example, 
%\beq
%\label{eq:uu}
%\avg{u_iu_j'} = {1\over3}\avg{u^2}\delta_{ij} - 
%{1\over2}\lt[{1\over2r}\diffr{}\, r^2\avg{\dul^2}
%\lt(\delta_{ij} - \rr_i\rr_j\rt) + 
%\avg{\dul^2}\rr_i\rr_j\rt],
%\eeq
%\beq
%\label{eq:uuu}
%\avg{u_iu_ju_k'} = - {1\over12}\lt[\avg{\dul^3}\delta_{ij}\rr_k 
%- {1\over2 r}\diffr{}\, r^2 \avg{\dul^3}\lt(\delta_{ik}\rr_j + \delta_{jk}\rr_i\rt)
%+ r^2\diffr{}{\avg{\dul^3}\over r}\,\rr_i\rr_j\rr_k\rt],
%\eeq
%where $\avg{\dul^2}$ and $\avg{\dul^3}$ 
%are the second- and third-order longitudinal structure functions of 
%the velocity field (see, e.g., \citealt{monin75}). 
%Taking into account isotropy, 
The evolution equation of the 
second-order correlation function of the velocity field 
is \citep{chandra51} 
\bea
\label{eq:vKH}
\difft{\avg{\ul\ul'}}=
{1\over r^4}\diffr{}\,r^4\lt(\avg{\ul^2\ul'} - \avg{\Bl^2\ul'}\rt) 
+ {2\nu\over r^4}\diffr{}r^4\diffr{\avg{\ul\ul'}} 
+ 2\avg{\fl\ul'}.
\eea
This is the von K\'arm\'an--Howarth equation for MHD turbulence. 
At $r=0$ it reduces to \eqref{eq:EK} 
for the evolution of the kinetic energy. For $r>0$, it contains 
information about the scale-by-scale energy budget (the turbulent 
cascade, energy exchanges between velocity and magnetic fields, etc.). 
If we now consider a statistically stationary state, in which all 
averages are constant in time, \eqref{eq:vKH} transforms into 
a direct generalisation of Komogorov's $4/5$ law for MHD 
turbulence:
\bea
\label{eq:ff}
\avg{\dul^3} - 6 \avg{\Bl^2\dul} 
- 6\nu\diffr{\avg{\dul^2}} = - {4\over5}\,\eps r, 
\eea
%In deriving this equation, 
%we have used \eqsdash{eq:uu}{eq:uuu}, 
%the vanishing of the one-point third-order correlator $\avg{\Bl^2\ul}=0$, and 
where we used the Taylor expansion $\avg{\fl\ul'}= (1/3)\epsilon + O(r^2/\lf^2)$, 
in which only the first term needs to be retained for $r\ll\lf$. 
%as long as point separations $r$ much smaller than the forcing scale $\lf$ 
%are considered. 
Note that the term involving mixed correlations of the velocity and magnetic 
fields contains the field $\vB$ rather than its increment $\dvB$ 
--- this hints at the nonlocality of the interaction between 
$\vu$ and $\vB$, which, as explained in \secref{sec:intro}, 
is the fundamental feature of MHD turbulence.

\Eqref{eq:ff} was derived by \citet{chandra51} 
(see also \citealt{polpouq98pre}). 
He also found the two other exact laws of MHD turbulence 
by constructing evolution equations analogous to \eqref{eq:vKH} 
for $\avg{B_iB_j'}$ and $\avg{u_iB_j'}$ 
(representing the magnetic-energy and the cross-helicity budgets, 
respectively). These laws relate certain 
third-order mixed correlation functions of $\vu$ and $\vB$ 
to each other and to the viscous and resistive dissipation. 
The laws can be written in 
several equivalent forms, none of which is physically illuminating 
in any obvious way. It is, therefore, as good a choice 
as any to cast them in a form that directly generalises results known 
in fluid turbulence in the way \eqref{eq:ff} does. 
This is achieved if we recall the alternative form of the equations 
of incompressible MHD in terms of the Elsasser variables $\vz^\pm=\vu\pm\vB$:
\bea
\difft{\vz^\pm} + \vz^\mp\cdot\vdel\vz^\pm = -\vdel p 
+ {\nu+\eta\over2}\Delta\vz^\pm + {\nu-\eta\over2}\Delta\vz^{\mp} + \vf,
\quad\vdel\vz^\pm=0.
\eea 
The form of the nonlinearity suggests a formal analogy with 
``passive'' advection of the field $\vz^+$ by the field $\vz^-$ 
and vice versa, and it is, therefore, not surprising that developing 
evolution equations for the correlators 
$\avg{z_i^\pm z_j^{\pm\prime}}$ leads to a pair of 
exact laws that are the MHD version of Yaglom's $4/3$ law for 
passive scalar \citep{polpouq98grl} 
\bea
\label{eq:ft}
\avg{\dzl^\mp|\dvz^\pm|^2} 
- \diffr{}\lt[(\nu+\eta)\avg{|\dvz^{\pm}|^2} + (\nu-\eta)\avg{\dvz^+\cdot\dvz^-}\rt]
= - {4\over3}\,\eps r, 
\eea
where $\delta\vz^\pm=\vz^\pm(\vx+\vr)-\vz^{\pm}(\vx)$.

\EMPTY{
There are two other exact laws, which can be derived in various ways. 
The more popular approach is based on the alternative form of the equations 
of incompressible MHD in terms of the Elsasser variables $\vz^\pm=\vu\pm\vB$:
\bea
\difft{\vz^\pm} + \vz^\mp\cdot\vdel\vz^\pm = -\vdel p 
+ {\nu+\eta\over2}\Delta\vz^\pm + {\nu-\eta\over2}\Delta\vz^{\mp} + \vf,
\quad\vdel\vz^\pm=0.
\eea 
The form of the nonlinearity suggests a formal analogy with 
``passive'' advection of the field $\vz^+$ by the field $\vz^-$ 
and vice versa, and it is, therefore, not surprising that developing 
evolution equations for the correlators 
$\avg{z_i^\pm z_j^{\pm\prime}}$ analogous to \eqref{eq:vKH} 
leads to a pair of exact laws that look like the MHD version of Yaglom's 
$4/3$ law for passive scalar \citep{polpouq98grl} 
\bea
\label{eq:ft}
\avg{\dzl^\mp|\dvz^\pm|^2} 
- \diffr{}\lt[(\nu+\eta)\avg{|\dvz^{\pm}|^2} + (\nu-\eta)\avg{\dvz^+\cdot\dvz^-}\rt]
= - {4\over3}\,\eps r. 
\eea
where $\delta\vz^\pm=\vz^\pm(\vx+\vr)-\vz^{\pm}(\vx)$.

Another approach is to construct evolution equations 
for $\avg{B_iB_j'}$ and $\avg{u_iB_j'}$, 
representing the magnetic-energy and the cross-helicity budgets, 
respectively \citep{chandra51}. The former requires the use of 
the induction equation \exref{eq:ind} only and gives in the 
statistically stationary state
\bea
\avg{\vu\cdot\dvB\Bl} - \avg{\ul\dvB\cdot\vB} = 
- {\eta\over 4}\diffr{\avg{|\dvB|^2}}
\eea
}

\section{Theoretical considerations: linear stretching and 
stripy fields \label{sec:theory}}

\subsection{Small-scale dynamo and its saturated state \label{sec:ssd}}

The small-scale dynamo is due to random stretching 
of the magnetic field by the velocity gradients associated 
with turbulence (\citealt{batch50,moffsaff64,kaz67}; 
see \citealt{schek06} for a review). 
In Kolmogorov turbulence, the velocity 
increments obey $\du(r)\sim(\eps r)^{1/3}$ 
for $\lf \gg r \gg \lvisc$, where 
$\lvisc\sim (\nu^3/\eps)^{1/4}\sim \lf\Re^{-3/4}$ is 
the viscous scale, $\Re=\urms\lf/\nu$ is the Reynolds number 
and $\urms=\avg{u^2}^{1/2}\sim (\eps\lf)^{1/3}$. 
Clearly, the viscous-scale motions give rise to the fastest stretching, 
with the rate $\gamma\sim\du(\lvisc)/\lvisc\sim (\eps/\nu)^{1/2} 
\sim (\urms/\lf)\Re^{1/2}$. 
The random stretching produces magnetic fields folded in long flux sheets 
with fold length $\lpar\sim\lvisc$ and 
direction reversals at the resistive scale
$\lres\sim(\eta/\gamma)^{1/2}\sim\lvisc\Pm^{-1/2}$. 
%(this is obtained by 
%balancing the stretching rate $\gamma$ and the resistive dissipation 
%rate $\eta/\lres^2$). 
When $\Pm\gg1$, the resistive scale $\lres$ is much smaller than the 
viscous scale $\lvisc$. 

While the folded fields are formally small-scale, they are 
capable of exerting a back reaction on the flow that has 
long-scale spatial coherence --- this is because the tension 
force in \eqref{eq:ind} is quadratic in $\vB$, so the direction 
reversals do not matter (note also that $\vB\cdot\vdel\vB\sim B^2/\lpar$, 
independent of $\lres$). \citet{batch50} proposed that the 
small-scale dynamo would saturate when the magnetic field is 
strong enough to oppose the stretching action of the viscous-scale 
motions, $\avg{B^2}\sim (\eps\nu)^{1/2}\sim\avg{u^2}\Re^{-1/2}$ 
(see also \citealt{moffatt63}). The alternative view expressed 
by \citet{schlbier50} was that 
the stretching action of the inertial-range motions would 
also be suppressed and $\avg{B^2}\sim\avg{u^2}$ in saturation. 
\citet{schek04} argued that the latter view is more consistent 
with numerical evidence but that, contrary to 
Schl\"uter and Biermann's assumption of scale-by-scale equipartition, 
the magnetic field in the saturated state remained 
small-scale-dominated and folded, with 
folds elongating to the forcing scale, $\lpar\sim\lf$. 
They proposed that saturation is controlled by the 
balance of the stretching of the field by the forcing-scale 
motions and the back reaction from the folded fields. 
The resistive scale is set by the characteristic 
time of the forcing-scale motions, 
$\lres\sim(\eta\lf/\urms)^{1/2}\sim\lf\Rm^{-1/2}$. 
This can still be much smaller than the viscous scale~$\lvisc$.  

%As we explained in \secref{sec:ssd}, when $\Pm\gg1$, 
%the resistive scale $\lres$ is much smaller than the viscous 
%scale $\lvisc$. 
Let us now examine \eqsand{eq:ff}{eq:ft} in the subviscous scale range, 
$r\ll\lvisc$. 

\subsection{The $4/5$ law at the subviscous scales \label{sec:ff_stokes}}

At subviscous scales, the velocity field is smooth, i.e., 
the velocity increments can be approximated by Taylor 
expansion: $\dvu \simeq \vr\cdot\vdel\vu(\vx)$, where 
$\vdel\vu$ does not depend on $\vr$. Let us substitute 
this into \eqref{eq:ff}. 
The term $\avg{\dul^3}\sim r^3$ is subdominant. 
\EMPTY{
\bea
\label{eq:BBu}
6\avg{\Bl^2\dul} &\simeq& 6 \avg{B_k B_l\nabla_i u_j} 
\avg{\rr_i\rr_j\rr_k\rr_l} r 
= {4\over 5}\avg{\vB\vB:\vdel\vu} r,\\
\label{eq:Sll}
\avg{\dul^2} &\simeq& \avg{\lt(\vrr\vrr:\vdel\vu\rt)^2}r^2
= \avg{\nabla_i u_j\nabla_k u_l} 
\avg{\rr_i\rr_j\rr_k\rr_l} r^2
= {1\over15}\avg{|\vdel\vu|^2}r^2,
\eea
where we have used 
$\avg{\rr_i\rr_j\rr_k\rr_l} = (1/15)\lt(\delta_{ij}\delta_{kl} 
+ \delta_{ik}\delta_{jl} + \delta_{il}\delta_{jk}\rt)$.
}
Since 
\bea
\label{eq:Sll}
\avg{\dul^2} \simeq \avg{\lt(\vrr\vrr:\vdel\vu\rt)^2}r^2
= \avg{\nabla_i u_j\nabla_k u_l} 
\avg{\rr_i\rr_j\rr_k\rr_l} r^2
= {1\over15}\avg{|\vdel\vu|^2}r^2,
\eea
the viscous term is $(4/5)\nu\avg{|\vdel\vu|^2}r$. 
The magnetic term is 
\bea
\label{eq:BBu}
6\avg{\Bl^2\dul} \simeq 6 \avg{B_k B_l\nabla_i u_j} 
\avg{\rr_i\rr_j\rr_k\rr_l} r 
= {4\over 5}\avg{\vB\vB:\vdel\vu} r.
\eea
In \eqsdash{eq:Sll}{eq:BBu}, we used 
$\avg{\rr_i\rr_j\rr_k\rr_l} = (1/15)\lt(\delta_{ij}\delta_{kl} 
+ \delta_{ik}\delta_{jl} + \delta_{il}\delta_{jk}\rt)$. 
We see that \eqref{eq:ff} reduces to the power balance 
\bea
\label{eq:powerbal}
\epsilon = \avg{\vB\vB:\vdel\vu} + \nu\avg{|\vdel\vu|^2},
\eea 
which is the steady-state form of \eqref{eq:EK}. Thus, the $4/5$ law 
carries no new information --- it simply 
registers the fact that the injected kinetic energy is 
partly dissipated viscously, partly transferred into the 
magnetic field. The latter part is dissipated resistively: 
$\avg{\vB\vB:\vdel\vu} = \eta\avg{|\vdel\vB|^2}$ from 
the steady-state form of \eqref{eq:EM}. 

\begin{figure}
\centering
\includegraphics[width=\vissize]{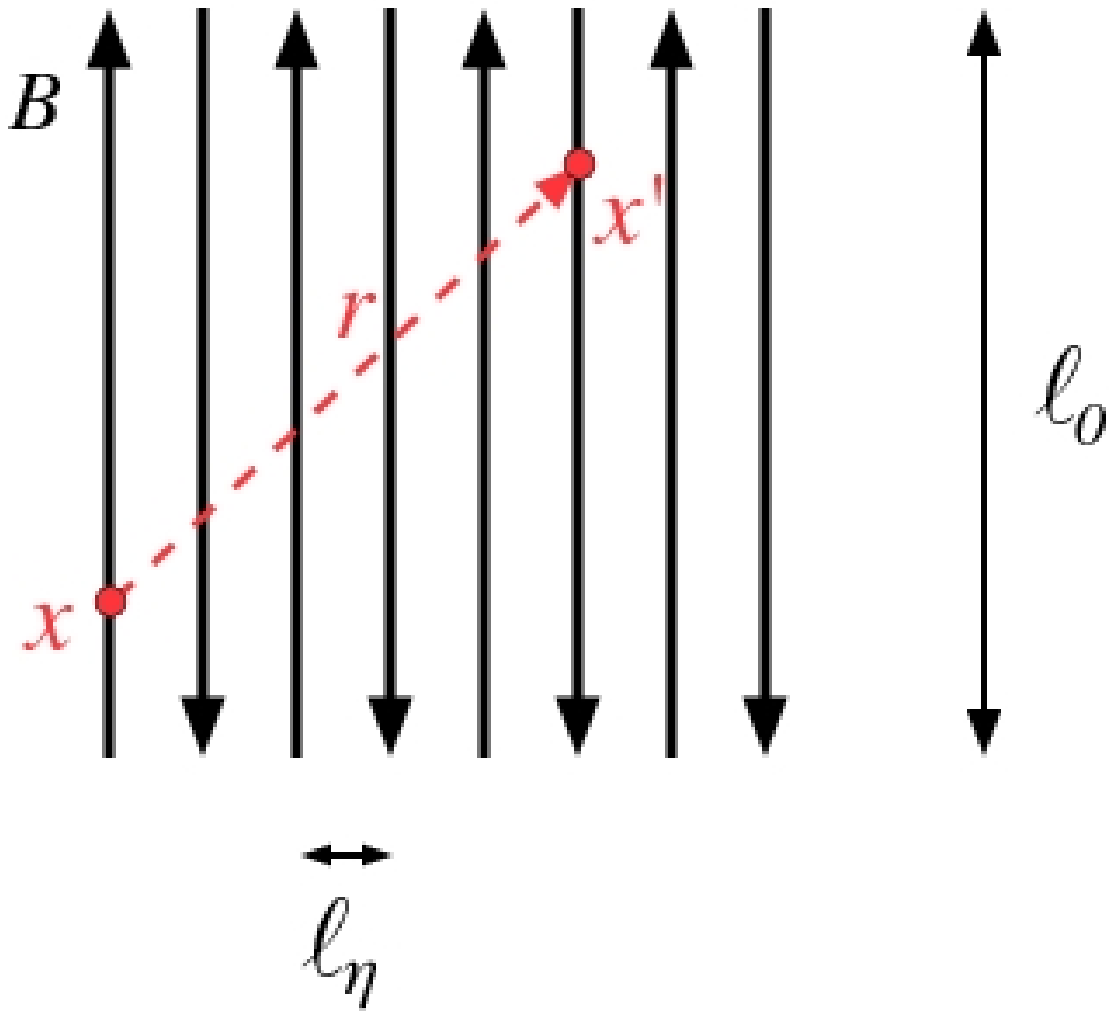}
\hskip10pt
\includegraphics[width=\vissize]{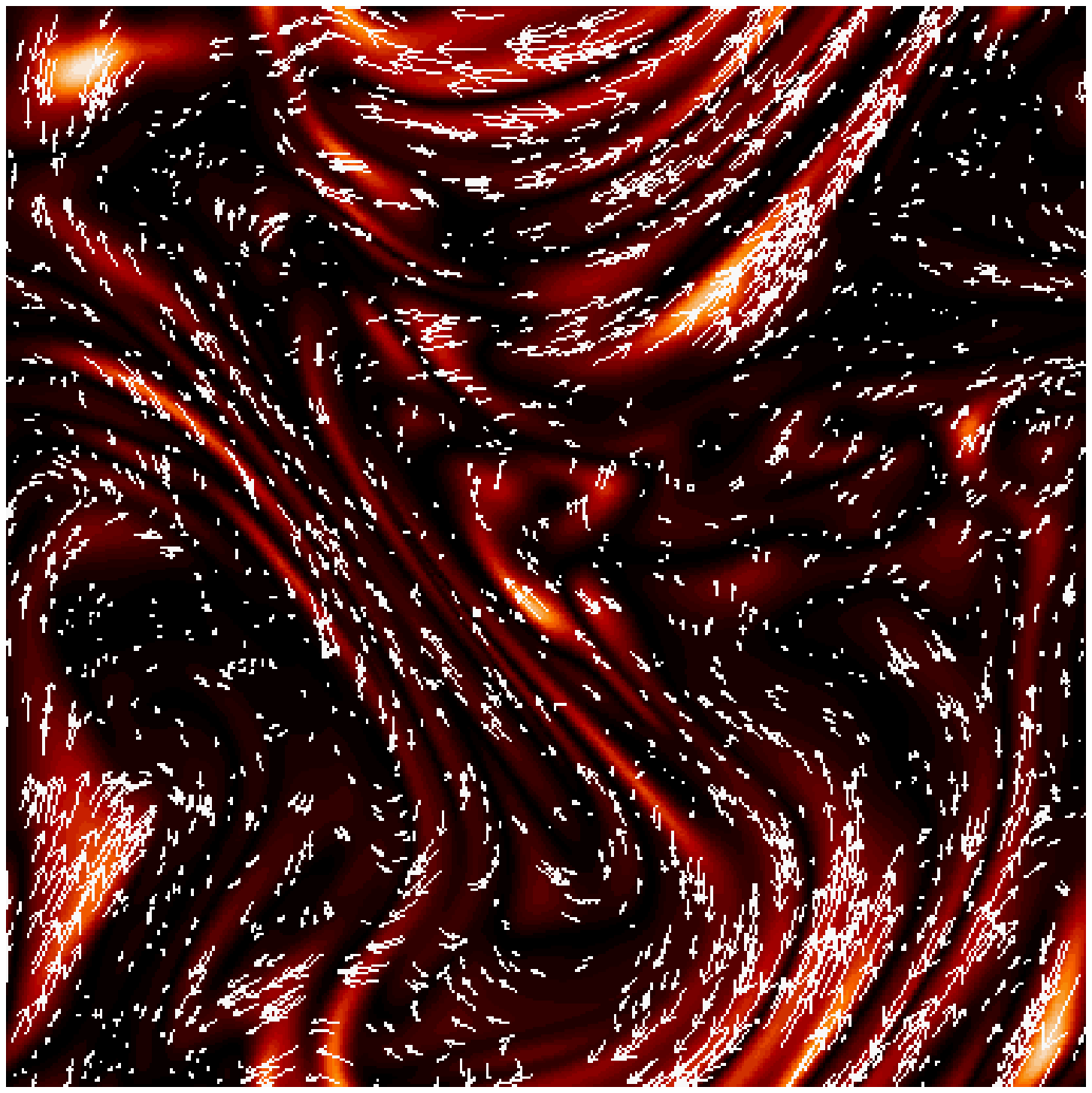}
\caption{({\em a}) The stripy field model. 
({\em b}) 
A cross section of $|\vB|$ for the run 
used in \figref{fig:stokes}{a} ($\Pm=1250$). 
Arrows indicate the in-plane direction of the field.
} 
\label{fig:stripy}
\end{figure}

\subsection{The $4/3$ law and the stripy-field model}

The $4/3$ law, \eqref{eq:ft}, is more interesting. 
In terms of $\dvu$ and $\dvB$, it~reads 
\bea
\nonumber
\avg{\dzl^\mp|\dvz^\pm|^2} 
&=& \avg{\dul|\dvu|^2} \pm 2 \avg{\dul\dvu\cdot\dvB} \mp \avg{\dBl|\dvu^2|}\\ 
\nonumber
&&+\,\, \avg{\dul|\dvB|^2} - 2\avg{\dBl\dvu\cdot\dvB} \mp \avg{\dBl|\dvB|^2}\\
&=& -{4\over3}\,\eps r 
+ 2\diffr{}\lt[\nu\avg{|\dvu|^2} \pm (\nu+\eta)\avg{\dvu\cdot\dvB} + \eta\avg{|\dvB|^2}\rt]. 
\label{eq:ftuB} 
\eea
Since $\dvu\sim r$, the first term on the left-hand side is $\sim r^3$, 
the second and third terms are $\sim r^2$, so all three are subdominant 
and can be dropped. The remaining terms depend on the structure of the 
magnetic field. Let us consider a drastically simplified model of this 
structure based on the understanding that the field is organised in 
folds with direction reversals at the resistive scale. 
The model is illustrated in \figref{fig:stripy}{a}. The orientation of
$\vB$ on this sketch must be understood as one among many
possible orientations with equal probability, since the flow is
statistically isotropic. The field is taken to be straight 
(or coherent along itself on scales comparable to the forcing scale $\lf$)
and to oscillate rapidly with spatial period $2\lres\ll\lf$ 
across itself. Given an arbitrary point $\vx$, 
if we randomly pick a point $\vx'$ on a sphere centered at $\vx$ 
with a fixed radius $r$ such that $\lres\ll r \ll\lvisc$, 
the field increment will be 
%\beq
%\begin{array}{llll}
%\dvB & = & -2\vB & \mbox{with probability 1/2},\\ 
%\dvB & = & 0     & \mbox{with probability 1/2}.
%\end{array}
%\eeq
$\dvB=-2\vB(\vx)$ with probability $1/2$ and $\dvB=0$ with probability $1/2$. 
We may now use this simple model and $\avg{\rr_i\rr_j}=(1/3)\delta_{ij}$ 
to calculate 
\bea
\label{eq:ulBB}
\avg{\dul|\dvB|^2} &=& 2\avg{B^2\nabla_i u_j}\avg{\rr_i\rr_j} r 
= {2\over3}\avg{B^2\vdel\cdot\vu} r = 0,\\
\label{eq:BluB}
2\avg{\dBl\dvu\cdot\dvB} &=& 4\avg{B_i B_j\nabla_k u_j}\avg{\rr_i\rr_k} r 
= {4\over3}\avg{\vB\vB:\vdel\vu} r,\\
\label{eq:BlBB}
\avg{\dBl|\dvB|^2} &=& -4\avg{B^2B_i}\avg{\rr_i} = 0,\\
%\eea
%and, for the second-order structure functions,
%\bea
\label{eq:uu}
\avg{|\dvu|^2} &=& \avg{\nabla_i u_j\nabla_k u_j}\avg{\rr_i\rr_k} r^2 
= {1\over 3}\avg{|\vdel\vu|^2} r^2,\\
\avg{\dvu\cdot\dvB} &=& - \avg{B_i\nabla_j u_i}\avg{\rr_j} r = 0,\\
\label{eq:BB}
\avg{|\dvB|^2} &=& 2\avg{B^2}. 
\eea
Substituting these expressions into \eqref{eq:ftuB}, we see that it reduces 
to the power balance \exref{eq:powerbal} just as the $4/5$ law did. 
It is essential, however, that to achieve this seemingly trivial outcome, 
we had to make assumptions about the spatial structure of the magnetic 
field. We have thus shown that a stripy alternating field is consistent 
with the $4/3$ law. 

It is perhaps appropriate to stress that the stripy field
sketched in \figref{fig:stripy}{a} is a highly idealised model 
of the field structure that does not take into account, 
for example, such features as turning points, bending of the folds 
on the scale of the flow, the extent to which the stripes are volume filling, 
possibility of some degree of misalignment of the 
alternating fields, etc. However, 
it has allowed us to give a particularly simple demonstration 
of the way the $4/3$ law accommodates the folded magnetic fields 
evident in \figref{fig:stripy}{b}. The key to understanding why 
such a simple model appears to be sufficient at least on the 
qualitative level may be that the fields 
best described by this model are also the strongest ones, so 
the model captures well the statistical quantities weighed towards 
the parts of the system where the field strength is largest. 
%a useful starting point for understanding the nature of the dynamo-generated field 

\section{Numerical Tests \label{sec:num}}

We now consider how \eqsand{eq:ff}{eq:ft} are satisfied in numerical 
experiments. 
%and what can be learned from that about the physics 
%of isotropic MHD turbulence. 
We use the data from direct numerical simulations of 
isotropic MHD turbulence by \citet{schek04}. 
The list of runs and other details are provided in that paper. 
These are three-dimensional spectral simulations in a periodic cube of size 1. 
They employ a white-noise random forcing at the 
box scale. The average power input is $\eps=1$. 

\subsection{The averaging procedure \label{sec:avg}} 

%The averaging procedure used here to calculate 
%the two-point correlation and structure functions is as follows. 
We use three levels of averaging to obtain two-point correlation 
and structure functions. 

\begin{itemize}

\item {\em Spherical average.} For a given point $\vx$ 
and a given point separation $r$, we average over all possible 
orientations of $\vec{r}$. We use 200 discrete orientations. 
In order to obtain accurate results, 
it is necessary to resort to linear interpolation between grid 
points to compute the fields at $\vx'=\vx+\vr$. 
Using the spherical average ensures that spurious anisotropies
introduced by finite size Cartesian grid effects %\citep{biferale00} 
are filtered out. It is not restrictive, because the turbulence 
is statistically isotropic. 

\item {\em Volume average.} Since the turbulence is spatially 
homogeneous, we average over all points $\vx$. In practice, 
the points $\vx$ are picked on a submesh that is 10 
times sparser than the mesh of the simulations. 

\item {\em Time average.} Finally, we average over about 20 
box-crossing times. This is equivalent to averaging over many realisations 
of the turbulence. A relatively large number of independent
snapshots is necessary in order to obtain converged
correlation functions of odd power 
(see \citealt{rincon06}). 
The averaged quantities we report are sufficiently converged 
(within a few percent for third-order quantities, 
which have the slowest convergence rate) for a qualitative 
understanding of the scale-by-scale budgets. 
We do not attempt any high-precision determination of scaling 
exponents from this data. 
%The averaged quantities we report are accurate (converged) roughly 
%within $10\%$, which is sufficient for a qualitative understanding 
%of the scale-by-scale budgets. 

\end{itemize}

Note that in computing the two-point statistical quantities, 
we may only use points $\vx'$ within a radius $r$ 
up to 1/4 of the box size from any given point $\vx$ 
because at larger distances spurious correlations arise due 
to the periodicity of the box.

\subsection{Numerical results in the viscosity-dominated limit \label{sec:ft}} 

%Let us now examine the numerical data we have at our disposal. 
It is not currently possible to have a resolved numerical simulation 
with $\Pm\gg1$ and $\Re\gg1$. However, one can study the subviscous 
fields in numerical simulations with $\Pm\gg1$ and $\Re\sim1$
\citep[the viscosity-dominated limit; see][]{schek04}. 
This means that the inertial range physics is sacrificed and 
we study the saturated state of a small-scale dynamo with $\Rm\gg1$ 
and a single-scale ($\sim\lf$) randomly forced flow. 
We note that in this setting, the difference between the 
\citet{batch50} and \citet{schlbier50} scenarios of saturation 
mentioned in \secref{sec:ssd} cannot be detected.  

\begin{figure}
\centering
\includegraphics[width=\maxsize]{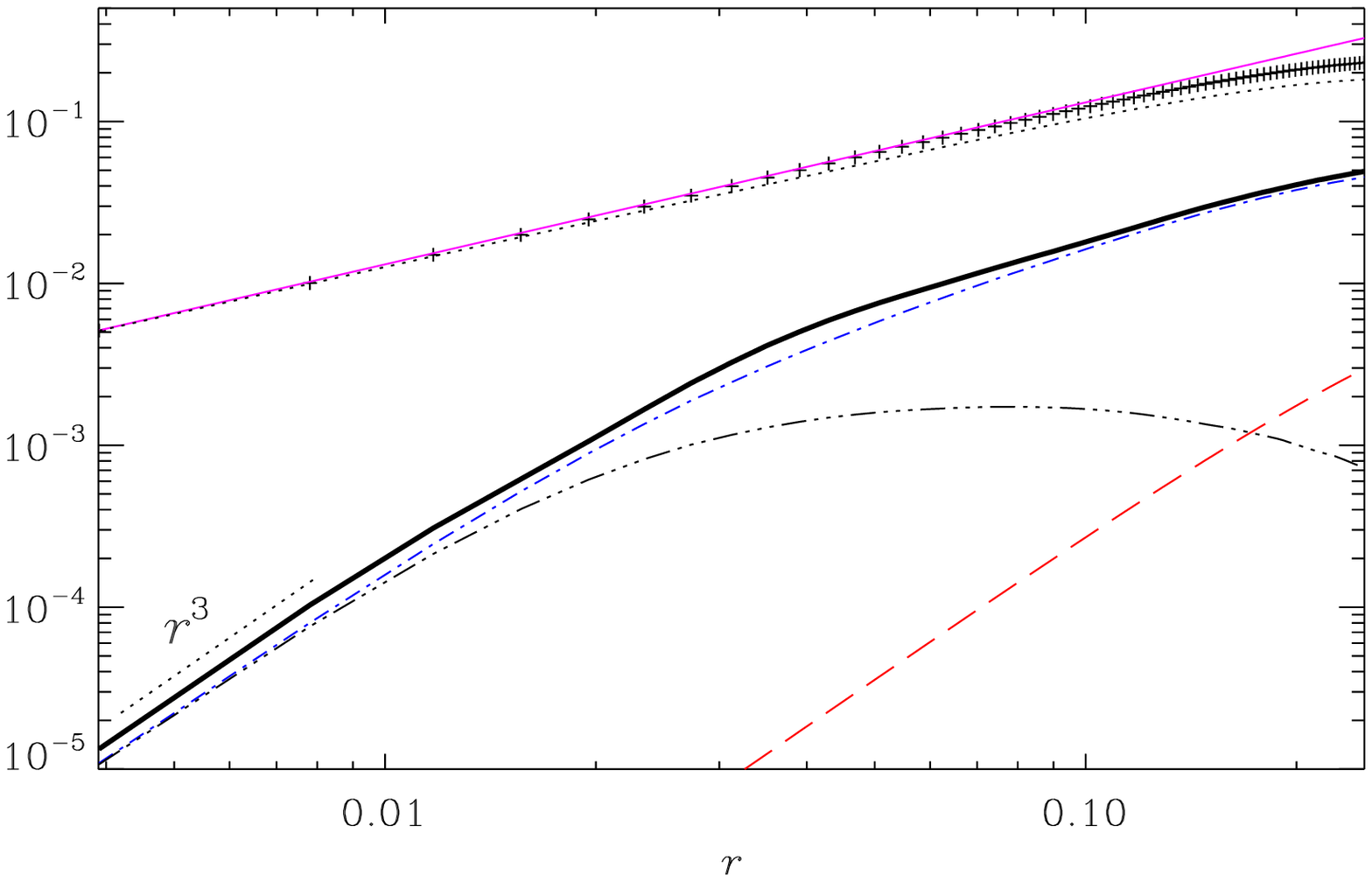}
\hfill
\includegraphics[width=\maxsize]{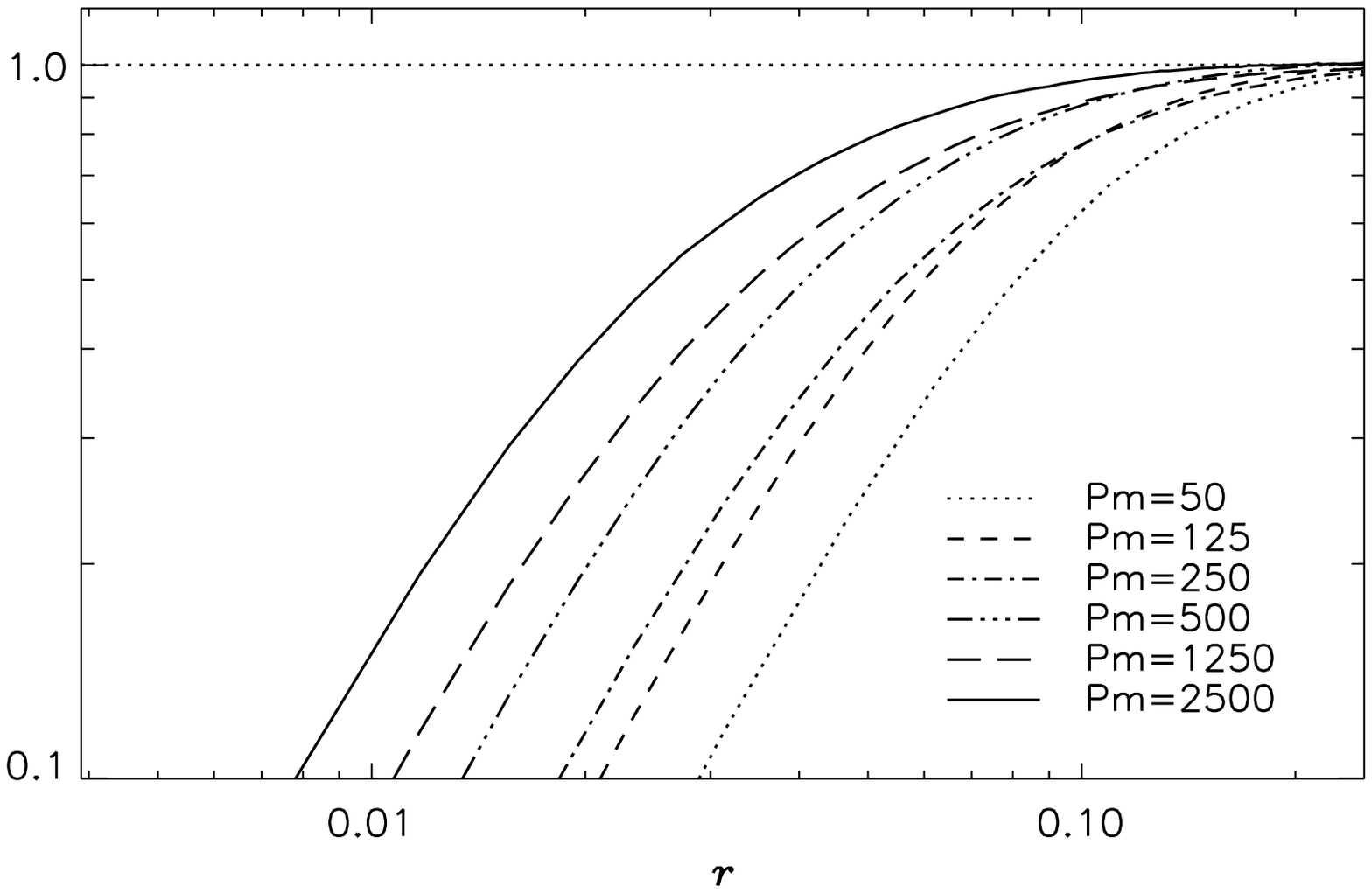} 
\caption{({\em a}) The $4/3$ law for a run 
with $\Pm=1250$, $\Re\sim1$ 
(run S5 of \citealt{schek04}: $256^3$, 
$\nu=5\times10^{-2}$, $\eta=4\times10^{-5}$). 
We plot minus the left- (\pluses) and the right-hand (\solid) sides 
of \eqref{eq:ft}, and also minus the dissipative term (\dotted). 
Several individual third-order structure functions in 
the left-hand side of \eqref{eq:ftuB} 
are shown: $-\avg{\dzl^-|\dvz|^2}$ (\boldsolid), 
$2\avg{\dBl\dvB\cdot\dvu}$ (\dashdotted),
$-\avg{\dul|\dvu|^2}$ (\dashed),
$-\avg{\dul|\dvB|^2}$ (\dashtridot). 
The three functions that are not shown are small. 
({\em b}) The second-order structure function $\avg{|\dvB|^2}/2\avg{B^2}$ 
for a sequence of runs with increasing $\Pm$ and 
constant $\Re\sim1$ (runs S1--S6 of \citealt{schek04}).} 
\label{fig:stokes} 
\end{figure}

Since the viscosity is large, the flow is smooth at all scales. 
As we explained in \secref{sec:ff_stokes}, the $4/5$ law 
in this regime does not contain any information beyond the power 
balance \exref{eq:powerbal}. Numerical results are in line 
with this expectation (not shown). 
The consistency of the $4/3$ law is illustrated in 
\figref{fig:stokes}{a}, which shows that the 
left- and right-hand sides of \eqref{eq:ft} independently 
computed from the numerical data match each other. 
The deviations at large scales 
are due to the higher-order terms in the Taylor expansion 
$\avg{\vf\cdot\vu'}=\epsilon + O(r^2/\lf^2)$. 
In \eqref{eq:ft}, the dominant term in the left-hand side 
is the viscous one. However, for a smooth velocity field, 
it is linear in $r$ (\eqref{eq:uu}), and, indeed, we find that the 
third-order structure function 
$-\avg{\dzl^-|\dvz^+|^2} \simeq (4/3)\lt(\eps-\nu\avg{|\vdel\vu|^2}\rt)r$
is also linear in $r$ for $r\gtrsim\lres$. 
The main contribution to it is due to $2\avg{\dBl\dvu\cdot\dvB}$, 
while all other terms in the left-hand side of \eqref{eq:ftuB} 
are subdominant. This is consistent with what is predicted 
by the stripy-field model, \eqsdash{eq:ulBB}{eq:BlBB}, 
although \eqref{eq:BluB} is only satisfied 
approximately (within $\sim30\%$). Note that for $r\ll\lres$, 
the magnetic field is smooth, $\dvB\sim r$, so the transition 
from the subresistive scaling $\avg{\dzl^-|\dvz^+|^2}\sim r^3$ 
to $\avg{\dzl^-|\dvz^+|^2}\sim r$ shows what the effective 
magnitude of the resistive scale is: $\lres\sim0.03$ for 
the run used in \figref{fig:stokes}{a}. 

\subsection{Scaling of the magnetic-field increments and 
the magnetic-energy spectrum} 

The stripy-field model implies that $\dB(r)\sim\Brms = \const$ 
for $\lres\ll r\ll \lvisc$ (\eqref{eq:BB}).  
Note that $\avg{|\dvB|^2}\to2\avg{B^2}$
in the limit $r\to\infty$ for any field structure, as long as the field 
decorrelates at long distances, $\avg{\vB'\cdot\vB}\to 0$. 
The nontrivial property of the stripy-field 
model is that this value is reached already at $r\gtrsim\lres$ 
because $\avg{\vB'\cdot\vB}=0$ due to direction reversals. 
\Figref{fig:stokes}{b} shows that the scale at which 
the value $\avg{|\dvB|^2}\simeq2\avg{B^2}$ is reached 
does indeed decrease as $\Rm$ is increased. 

What the magnetic-energy spectrum is in the saturated state 
is not currently known. 
The stripy-field result $\dB\sim r^0$ appears to imply that 
the spectrum is $k^{-1}$. 
However, the stripy-field model does not take into account 
that the alternating fields do not cancel each other 
perfectly, so there should be a gradual loss of correlation 
as $r$ increases. If this effect is important, the correlation 
function $\avg{|\dvB|^2}$ may have a peak around $r\sim\lres$ 
followed by a gradual fall off to the value $2\avg{B^2}$ at $r\gg\lres$.
If this fall off is a power law $\sim (\lres/r)^{1+\alpha}$, 
the spectrum should then be $k^\alpha$. 
For example, in the kinematic regime of the small-scale dynamo, 
theory predicts $\alpha=3/2$ \citep{kaz67}. 
In the saturated state,
the numerically computed correlation functions (\figref{fig:stokes}{b})
show no sign of having a peak, which suggests $\alpha=-1$. 
It cannot, however, be ruled out that, as suggested in \citet{schek04}, 
the spectrum is flatter, possibly even $\alpha>0$, but the convergence 
to the asymptotic scalings is slow and the values of $\Pm$ 
used in our simulations are insufficient to determine $\alpha$ 
accurately. 

\begin{figure}
\centering
\includegraphics[width=\maxsize]{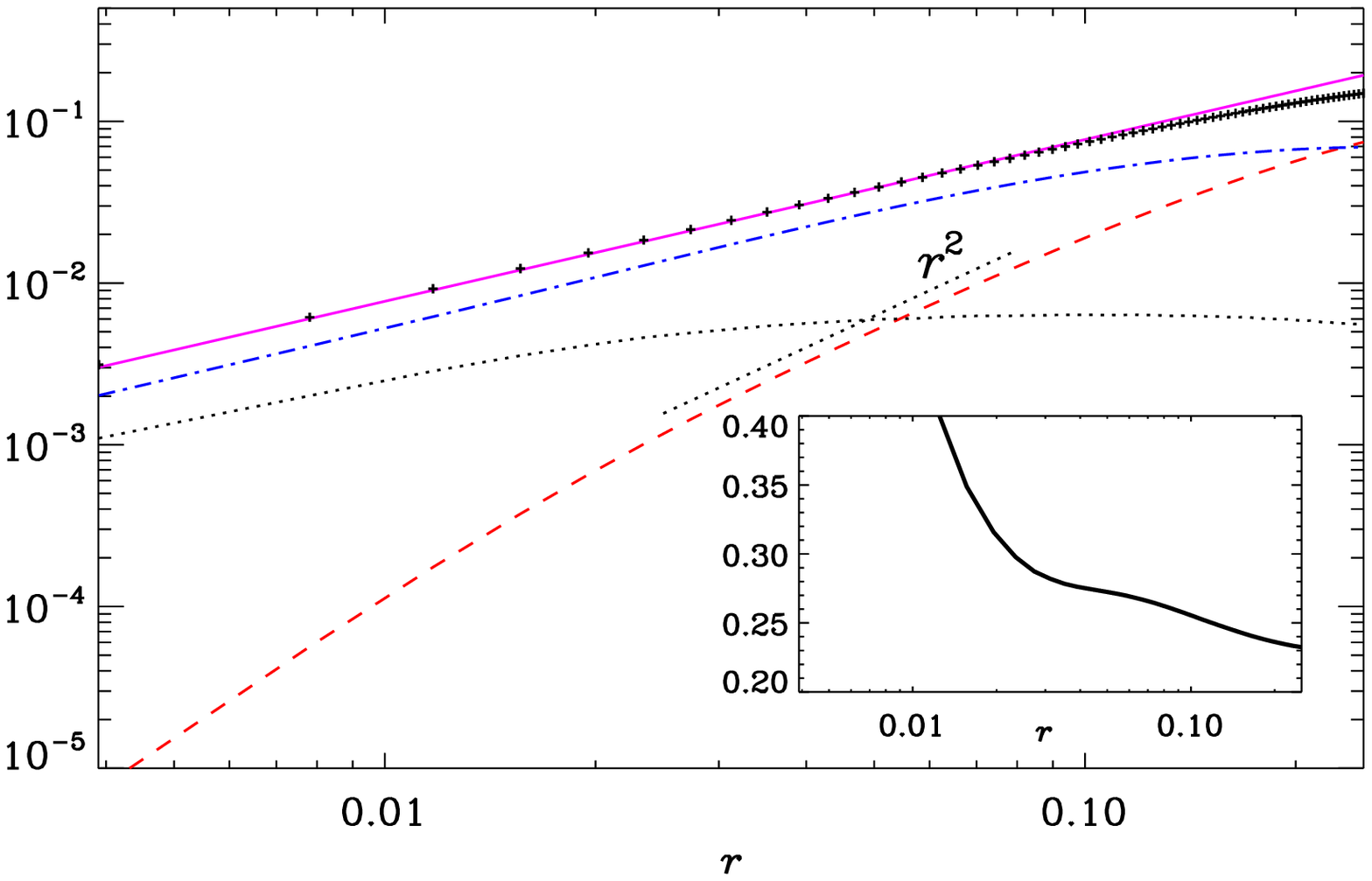}
\hfill
\includegraphics[width=\maxsize]{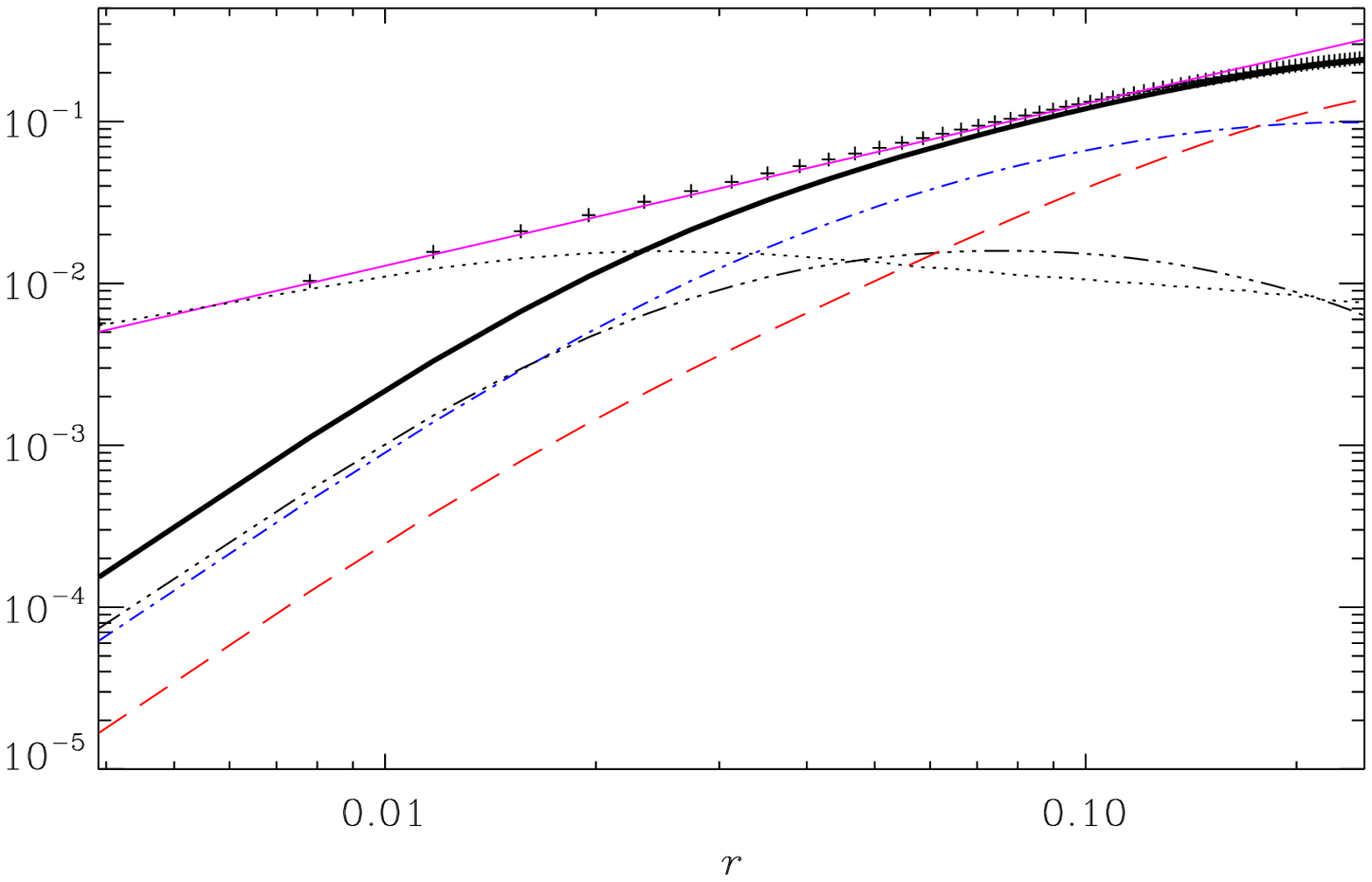}
\caption{
({\em a}) 
The $4/5$ law for a run 
with $\Pm=1$, $\Re\simeq 400$ (run A of \citealt{schek04}: $256^3$, 
$\nu=\eta=5\times10^{-4}$). 
We plot minus the left- (\pluses) and the right-hand (\solid) sides 
of \eqref{eq:ff}, minus the viscous term (\dotted), 
$-\avg{\dul^3}$ (\dashed) and $6\avg{\Bl^2\dul}$ (\dashdotted).
Inset: plot of $\lstar = -6\avg{\Bl^2\dul}r/\avg{\dul^3}$. 
({\em b}) The $4/3$ law for the same run. 
See caption of \figref{fig:stokes}{a} for explanation of line styles. 
The functions $\avg{\dul\dvu\cdot\dvB}$, $\avg{\dBl|\dvu|^2}$ and 
$\avg{\dBl|\dvB|^2}$ are again small (not shown).} 
\label{fig:runA}
\end{figure}

\subsection{Numerical results for $\Pm=1$}

While the limit $\Rm\gg\Re\gg1$ is unresolvable, 
some idea of the structure of isotropic MHD turbulence in a regime that is 
not viscosity-dominated can be gained by studying the case of $\Pm=1$, 
which has been the favourite choice in numerical experiments 
\citep[e.g.,][]{haugen04}. Such simulations may, in fact, 
be more relevant than it would appear to understanding the $\Pm\gg1$ regime 
if the saturated state of the small-scale dynamo is controlled by the 
interaction of the folded magnetic fields with the forcing-scale 
velocities (\secref{sec:ssd}), rather than --- 
as \citet{batch50} thought --- with the viscous-scale ones. 
%(see \citealt{schek06} for a historical review). 

Whether this is true is best addressed numerically by comparing the four 
terms in the $4/5$ law, \eqref{eq:ff}. This is done in \figref{fig:runA}{a} 
for a run with $\Re=\Rm\simeq 400$. The $4/5$ law is well satisfied: 
the independently computed left- and right-hand sides of \eqref{eq:ff} match
except for the large-scale discrepancy due to the higher-order 
contributions from the forcing term. 
The viscous term is subdominant for $r\gtrsim\lvisc\simeq0.05$
leaving some room for a possible inertial range. 
Except at $r\sim\lf$, the 
dominant balance is between %the magnetic-interaction term 
$6\avg{\Bl^2\dul}$ and $(4/5)\eps r$, while $\avg{\dul^3}$ is 
subdominant. Physically this means that {\em the kinetic-energy 
cascade is short-circuited by the energy transfer into the magnetic field.}
Interestingly, for the subdominant ``cascade'' term, we~find empirically 
\bea
-\avg{\dul^3}\simeq 6\avg{\Bl^2\dul} {r\over\lstar} 
\simeq {4\over5}{\eps\over\lstar}\,r^2,
\label{eq:du3_scaling}
\eea
where $\lstar\simeq0.25$ is some typical length, 
comparable to the forcing scale $\lf$. The resolution is limited 
and the scale range where $\avg{\dul^3}\propto r^2$ 
holds is not wide. Numerically, the relation 
$\avg{\dul^3}\propto\avg{\Bl^2\dul} r$ turns out to be satisfied 
across a somewhat wider interval (see inset in \figref{fig:runA}{a}). 
\Eqref{eq:du3_scaling} suggests that the velocity increments 
have the scaling $\du\sim r^{2/3}$. Indeed, 
the second-order structure function of $\vu$ computed 
from our data is 
consistent with $\avg{\dul^2}\sim r^{4/3}$ 
(not shown) and the kinetic-energy spectrum does appear to 
have the corresponding scaling $k^{-7/3}$, as noticed by \citet{schek04}. 

The $4/3$ law is well satisfied (\figref{fig:runA}{b}), 
as it was in the viscosity-dominated case 
(\secref{sec:ft}), but now there is an interval ($r\gtrsim 0.02$) 
where the dissipative terms are small and \eqref{eq:ft} 
becomes $-\avg{\dzl^-|\dvz^+|^2}\simeq(4/3)\eps r$. The 
numerically computed third-order 
structure function in the right-hand side of this equation 
does, indeed, scale linearly with $r$. The main contribution to 
this function still comes from 
$2\avg{\dBl\dvu\cdot\dvB}$ (see \eqref{eq:ftuB}), except at the 
largest scales, where $\avg{\dul|\dvu|^2}$ becomes important. 
That the $4/3$ law approximately reduces to a balance between 
the magnetic-interaction term $2\avg{\dBl\dvu\cdot\dvB}$ 
and $(4/3)\eps r$ with the kinetic-energy ``cascade'' term 
$\avg{\dul|\dvu|^2}$ interfering only at $r\sim\lf$ is 
in line with what we learned above from the 
$4/5$ law. The qualitative similarity of the internal 
structure of $\avg{\dzl^-|\dvz^+|^2}$ (the relative importance 
of the six terms in the left-hand side of \eqref{eq:ftuB}) to 
the viscosity-dominated case appears to suggest that the 
magnetic field in the $\Pm=1$ case still has a stripy (folded) 
structure. This is, indeed, supported by various quantitative 
tests \citep{schek04}. 
%as well as by examining visualisations 
%of the magnetic field for $\Pm=1$ runs (\ffigref{fig:runAsnaps}) 
%and comparing with the viscosity-dominated case (\figref{fig:stokes}{a}). 

%\begin{figure}
%\centering
%\includegraphics[width=\vissize]{uu_z583_zsh185_010_z.ps}
%\hskip10pt
%\includegraphics[width=\vissize]{BB_z583_zsh185_010_z.ps}
%\caption{
%Cross sections of $|\vu|$ ({\em a}) and $|\vB|$ ({\em b}) 
%for the run used in \ffigref{fig:runA} ($\Pm=1$). Arrows indicate the 
%in-plane direction of the field.
%} 
%\label{fig:runAsnaps}
%\end{figure}

\section{Discussion: turbulence in the presence of stripy field \label{sec:inertial}}

Numerical results appear to make a fairly compelling case that 
the structure of the magnetic field in the isotropic MHD turbulence 
is dominated by folds (stripes). 
We have demonstrated above that such a structure 
is consistent with the exact scaling laws. 
It has proven to be much more difficult 
to understand the structure of the velocity field. 
It is clear that the cascade of kinetic energy is short-circuited 
at the forcing scale, with injected energy diverted into 
maintaining the folds against continuous Ohmic dissipation. 
It is not clear from the available numerical results 
whether most or only a finite fraction of the injected power 
$\epsilon$ is thus diverted. If the latter is the case, 
i.e., if the magnetic-interaction term $6\avg{\Bl^2\dul}$ only 
cancels a part of the flux term $(4/5)\eps r$ in \eqref{eq:ff}, 
there should still be a kinetic-energy cascade from the forcing to 
viscous scale and it is hard to see from \eqref{eq:ff}
how the velocity increments associated with this cascade can fail to have 
the Kolmogorov scaling, $\avg{\dul^3}\simeq-(4/5)\chi\eps r$, 
where $\chi<1$ is some finite positive number.  
These motions are likely to be 
Alfv\'enic perturbations of the folded structure \citep{schek02}. 
If, on the other hand, the short-circuiting of the kinetic-energy 
cascade is nearly complete, i.e., $6\avg{\Bl^2\dul}\simeq(4/5)\eps r$, 
then a subdominant scaling is possible: $\avg{\dul^3}\sim r^\beta$, $\beta>1$. 
\Eqref{eq:du3_scaling}, which we deduced from our numerical experiment, 
appears to support this possibility with $\beta=2$. However, 
the resolution of our study was not sufficient to have 
a convincing parameter scan in $\Re$ and prove that the 
steeper-than-Kolmogorov scaling of the velocity increments 
is not simply due to too much viscosity. 
Note that \citet{haugen04} report kinetic-energy spectra 
that may be consistent with $k^{-5/3}$ at values of $\Re$ 
roughly twice as large as ours. 

We remark in passing that the short-circuiting of the 
kinetic-energy cascade appears also to be a feature of the 
turbulence in polymer solutions \citep{deangelis05,berti06}, 
the mathematical description of which is similar to MHD, 
with elastic polymer chains stretched by turbulence in 
the same way magnetic fields are. 

While the above study does not constitute the complete solution of the 
problem of isotropic MHD turbulence, it adds to the weight of evidence 
that {\em this turbulence is controlled by the direct nonlocal 
interaction between the forcing-scale motions and small-scale 
stripy, or folded, magnetic structures.} This nonlocality means that 
scaling arguments cannot be made along the usual, Kolmogorov-like, 
lines. This makes the problem conceptually difficult. %and hard to handle. 
The %modern remedy of choice in such cases, 
numerical simulations, while helpful, are not as yet 
sufficiently large to access the asymptotic regime of 
interest: ideally, $\Rm\gg\Re\gg1$. 
%Thus, one cannot adopt 
%a purely ``experimentalist'' approach and simply report 
%careful data analysis --- a measure of guesswork 
%and plausible reasoning is necessary both in thinking about 
%the physics and in interpreting numerical results. 
The exact scaling laws provide the only 
rigorous quantitative constraints available in the theory 
of turbulence and are, therefore, useful. 
In this paper, we have attempted to discern the message that 
these laws carry about the structure of isotropic MHD turbulence. 
Given the limited resolution of the numerical experiments that 
we used to guide us, it may be fruitful to return to this 
type of analysis when larger computations become feasible. 

\begin{acknowledgments}
We are grateful to S.\ C.\ Cowley 
and J.\ C.\ McWilliams for inspiring discussions. 
We also thank N.\ Haugen and A.~Brandenburg who were 
present at the inception of this project. 
T.Y.\ was supported by Crighton and UKAFF Fellowships 
and the Newton Trust, 
F.R.\ by the Leverhulme and Newton Trusts, 
A.S.\ by a PPARC Advanced Fellowship and King's College, Cambridge. 
Simulations were done at UKAFF (Leicester) and NCSA (Illinois). 
\end{acknowledgments}

%\label{lastpage}

\bibliographystyle{jfm}
\bibliography{yrs_JFM}

\begin{thebibliography}{18}
\expandafter\ifx\csname natexlab\endcsname\relax\def\natexlab#1{#1}\fi

\bibitem[Alexakis {\em et~al.\/}(2005)Alexakis, Mininni \&
  Pouquet]{alexakis05sat}
{\sc Alexakis, A., Mininni, P.~D. \& Pouquet, A.} 2005 Shell-to-shell energy
  transfer in magnetohydrodynamics. {I}. {S}teady state turbulence. {\em Phys.\
  Rev.~E\/} {\bf 72}, 046301.

\bibitem[Batchelor(1950)]{batch50}
{\sc Batchelor, G.~K.} 1950 On the spontaneous magnetic field in a conducting
  liquid in turbulent motion. {\em Proc.\ R.\ Soc.\ London, Ser.~A\/} {\bf
  201}, 405--416.

\bibitem[Berti {\em et~al.\/}(2006)Berti, Bistagnino, Boffetta, Celani \&
  Musacchio]{berti06}
{\sc Berti, S., Bistagnino, A., Boffetta, G., Celani, A. \& Musacchio, S.} 2006
  Small scale statistics of viscoelatic turbulence. {\em Europhys.\ Lett., {\rm
  submitted (e-print nlin.CD/0606043)}\!\!\/} .

\bibitem[Biskamp \& M\"uller(2000)]{bisk00}
{\sc Biskamp, D. \& M\"uller, W.-C.} 2000 Scaling properties of isotropic
  three-dimensional magnetohydrodynamic turbulence. {\em Phys.\ Plasmas\/} {\bf
  7}, 4889--4900.

\bibitem[Brandenburg(2001)]{brand01}
{\sc Brandenburg, A.} 2001 The inverse cascade and nonlinear alpha-effect in
  simulations of isotropic helical hydromagnetic turbulence. {\em
  Astrophys.~J.\/} {\bf 550}, 824--840.

\bibitem[Chandrasekhar(1951)]{chandra51}
{\sc Chandrasekhar, S.} 1951 The invariant theory of isotropic turbulence in
  magneto-hydrodynamics. {\em Proc.\ R.\ Soc.\ London, Ser.~A\/} {\bf 204},
  435--449.

\bibitem[De{\ }Angelis {\em et~al.\/}(2005)De{\ }Angelis, Cassciola, Benzi \&
  Piva]{deangelis05}
{\sc De{\ }Angelis, E., Cassciola, C.~M., Benzi, R. \& Piva, R.} 2005
  Homogeneous isotropic turbulence in dilute polymers. {\em J.~Fluid Mech.\/}
  {\bf 531}, 1--10.

\bibitem[Haugen {\em et~al.\/}(2004)Haugen, Brandenburg \& Dobler]{haugen04}
{\sc Haugen, N.~E.~L., Brandenburg, A. \& Dobler, W.} 2004 Simulations of
  nonhelical hydromagnetic turbulence. {\em Phys.\ Rev.~E\/} {\bf 70}, 016308.

\bibitem[Kazantsev(1968)]{kaz67}
{\sc Kazantsev, A.~P.} 1968 Enhancement of a magnetic field by a conducting
  fluid. {\em Sov.\ Phys.\ JETP\/} {\bf 26}, 1031--1034.

\bibitem[Moffatt(1963)]{moffatt63}
{\sc Moffatt, H.~K.} 1963 Magnetic eddies in an incompressible viscous fluid of
  high electrical conductivity. {\em J.~Fluid Mech.\/} {\bf 17}, 225--239.

\bibitem[Moffatt \& Saffman(1964)]{moffsaff64}
{\sc Moffatt, H.~K. \& Saffman, P.~G.} 1964 Comment on "{G}rowth of a weak
  magnetic field in a turbulent conducting fluid with large magnetic {P}randtl
  number". {\em Phys.\ Fluids\/} {\bf 7}, 155.

\bibitem[Politano \& Pouquet(1998{\natexlab{{\em a\/}}})]{polpouq98grl}
{\sc Politano, H. \& Pouquet, A.} 1998{\natexlab{{\em a\/}}} Dynamical length
  scales for turbulent magnetized flows. {\em Geophys.\ Res.\ Lett.\/} {\bf
  25}, 273--27.

\bibitem[Politano \& Pouquet(1998{\natexlab{{\em b\/}}})]{polpouq98pre}
{\sc Politano, H. \& Pouquet, A.} 1998{\natexlab{{\em b\/}}} {v}on
  {K}\'arm\'an-{H}owarth equation for magnetohydrodynamics and its consequences
  on third-order longitudinal structure and correlation functions. {\em Phys.\
  Rev.~E\/} {\bf 57}, R21--R24.

\bibitem[Rincon(2006)]{rincon06}
{\sc Rincon, F.} 2006 Anisotropy, inhomogeneity and inertial range scalings in
  turbulent convection. {\em J.~Fluid Mech.\/} {\bf 563}, 43--69.

\bibitem[Schekochihin \& Cowley(2006)]{schek06}
{\sc Schekochihin, A.~A. \& Cowley, S.~C.} 2006 Turbulence and magnetic fields
  in astrophysical plasmas. In {\em Magnetohydrodynamics: Historical Evolution
  and Trends\/} (ed. S.~Molokov, R.~Moreau \& H.~K. Moffatt). Berlin: Springer,
  in press (e-print astro-ph/0507686).

\bibitem[Schekochihin {\em et~al.\/}(2002)Schekochihin, Cowley, Hammett, Maron
  \& McWilliams]{schek02}
{\sc Schekochihin, A.~A., Cowley, S.~C., Hammett, G.~W., Maron, J.~L. \&
  McWilliams, J.~C.} 2002 A model of nonlinear evolution and saturation of the
  turbulent {MHD} dynamo. {\em New J.~Phys.\/} {\bf 4}, 84.

\bibitem[Schekochihin {\em et~al.\/}(2004)Schekochihin, Cowley, Taylor, Maron
  \& McWilliams]{schek04}
{\sc Schekochihin, A.~A., Cowley, S.~C., Taylor, S.~F., Maron, J.~L. \&
  McWilliams, J.~C.} 2004 Simulations of the small-scale turbulent dynamo. {\em
  Astrophys.~J.\/} {\bf 612}, 276--307.

\bibitem[Schl{\"u}ter \& Biermann(1950)]{schlbier50}
{\sc Schl{\"u}ter, A. \& Biermann, L.} 1950 Interstellare magnetfelder. {\em
  Z.~Naturforsch.\/} {\bf 5a}, 237--351.

\end{thebibliography}

\end{document}